\documentstyle[preprint,aps,epsf]{revtex}
\begin{document}
\tightenlines

\draft 

\author{C. Callegari, A. Conjusteau, I. Reinhard, K. K. Lehmann, and
  G. Scoles}

\address{Department of Chemistry, Princeton University, Princeton NJ
  08544, USA}

\author{F. Dalfovo}

\address{Dipartimento di Fisica, Universit\`a di Trento, I-38050 Povo,
  Italy \\
and Istituto Nazionale per la Fisica della Materia, Unit\`a di Trento.}

\date{\today}

\title{A Superfluid Hydrodynamic Model for the Enhanced Moments of Inertia of
Molecules in Liquid $^4$He}

\maketitle

\begin{abstract}
We present a superfluid hydrodynamic model for the increase in moment
of inertia, $\Delta I$, of molecules rotating in liquid $^4$He.  The
static inhomogeneous He density around each molecule (calculated using
the Orsay--Paris liquid $^4$He density functional) is assumed to
adiabatically follow the rotation of the molecule.  We find that the
$\Delta I$ values created by the viscousless and irrotational flow are
in good agreement with the observed increases for several molecules
\protect{[}OCS, (HCN)$_2$, HCCCN, and HCCCH$_3$\protect{]}.  For HCN
and HCCH, our model substantially overestimates $\Delta I$. This is
likely to result from a (partial) breakdown of the adiabatic following
approximation.
\end{abstract}

The spectroscopy of atoms and molecules dissolved in He nanodroplets
provides both a new way to study microscopic dynamics of this unique
quantum fluid~\cite{greb98}, and a very cold matrix
(0.4~K~\cite{hart95}) to create and study novel
species~\cite{higg96,lehm98,nauta99c}.  Recent experiments have
demonstrated that even heavy and anisotropic molecules display
rotationally resolved vibrational spectra with a structure reflecting
the gas phase symmetry of the molecule.  However, the rotational
constants required to reproduce the spectra are often substantially
reduced from those of the isolated molecule.  For example, the
${\rm \nu_3}$ vibrational band of SF$_6$ dissolved in He
nanodroplets (first observed by Goyal {\it et al.}~\cite{goya92} and
later rotationally resolved and analyzed by Hartmann {\em et
al.}~\cite{hart95,froc94}) indicates that the effective moment of
inertia, $I_{\rm eff}$, in liquid $^4$He is 2.8 times that of the
isolated molecule.  The same qualitative behavior has been found for a
wide range of other molecules~\cite{greb98,nauta99a,conj99}.  In an
elegant recent experiment, it has been demonstrated that the
rotational structure of OCS broadens and collapses in pure $^3$He
droplets, and is recovered when $\approx 60$ $^4$He atoms are
co-dissolved in the $^3$He~\cite{greb98}.  The association of the
weakly damped, unhindered rotation with the Bose symmetry of $^4$He
suggests that this phenomenon is a manifestation of superfluidity, and
has been called the microscopic Andronikashvili
experiment~\cite{greb98}.

A theory able to reproduce the observed increase, $\Delta I$, in
molecular moments of inertia would be of interest for at least two
reasons.  First, the enhanced inertia provides a window into the
dynamics of the liquid.  Second, the ability to predict the rotational
constants would further improve the utility of He nanodroplet
isolation spectroscopy for the characterization of novel chemical
species.

The first model proposed to explain the observed $\Delta I$ assumed
that a certain number of He atoms, trapped in the interaction
potential of the solute, rotate rigidly with the latter~\cite{hart95}.
In the case of SF$_6$, 8 He atoms trapped in the octahedral global
potential minima would create a rigidly rotating `supermolecule' that
would have approximately the observed $I_{\rm eff}$.  In the case of
OCS, putting a six He atom `donut' in the potential well around the
molecule also reproduces the observed $I_{\rm eff}$~\cite{greb99}.  Recent
Diffusion Monte Carlo (DMC) calculations~\cite{lee99} have predicted
that the effective rotational constant of SF$_6$--He$_N$
monotonically decreases from that of the isolated molecule to the
large cluster limit, reached at $N=8$, and remains essentially
constant for $N = 8$--$20$.
The supermolecule model has been recently extended to consider the
rigid rotation of a `normal fluid fraction' of the He density which is
claimed to be significant only in the first solvation
layer~\cite{greb98,greb99}, based on Path Integral Monte Carlo
calculations of Kwon {\em et al.} \cite{kwon96} which show a
molecule-induced reduction of the superfluid fraction. These
calculations have been recently used to propose a definition of a
spatially dependent normal fluid fraction which reproduces the
observed $I_{\rm eff}$ of solvated SF$_6$~\cite{kwon99}.

The limitations of the supermolecule model are made clear by the
$\Delta I$ observed for HCN in He droplets~\cite[a]{nauta99a} which is
only $\approx 5\%$ of the $\Delta I$ observed upon formation of a gas
phase He$\cdot$HCN van der Waals complex~\cite{druc95}.  Furthermore,
it has been previously recognized that in principle there is also a
superfluid hydrodynamic contribution, $I_{{\rm h}}$, to $\Delta
I$~\cite{greb98,whaley98,lehm99}.  Previous estimates, based upon a
classic treatment of the rotation of an ellipsoid in a fluid of
uniform density, found that $I_{{\rm h}}$ is only a small fraction
of the observed $\Delta I$, at least for heavy rotors such as
OCS~\cite{greb99,lehm99}.  In this report, we show that if the spatial
variation of the He solvation density around the solute molecule is
taken into account, the calculated $I_{{\rm h}}$ is instead rather
large and agrees well with experimental data.  We compare
our calculations with the experimental results available in
the literature (OCS \cite{greb98}, HCN \cite[a]{nauta99a}) and with
results recently obtained in our laboratory for HCCCH$_3$ and HCCCN,
and in the laboratory of R.~E.~Miller for (HCN)$_2$~\cite[b]{nauta99a}
and HCCH~\cite[c]{nauta99a}.

We first calculate the ground state He density, $\rho$, around a
static solute molecule.  The molecule is then considered to undergo
classical rotation, slowly enough that the helium ground state density
adiabatically follows the molecular rotation.  The kinetic energy
associated with the He flow (assumed viscousless and irrotational)
is used to calculate $I_{{\rm h}}$.

The main input of our hydrodynamic model is the ground state density of He
around the solute.  DMC calculations can provide this density with a
minimum of assumptions beyond the interaction
potentials~\cite{barn93}, but are computationally expensive.  The
Density Functional method, which is a good compromise between accuracy
and computational cost, consists in numerically minimizing the total
energy of the many-body system in the form of a semi-empirical
functional of the He density: $E=\int d{\bf r}\, {\cal H}[\rho({\bf
r})]$. The energy density ${\cal H}$ contains an effective non-local
interaction with a few parameters fixed to reproduce known properties
of bulk liquid He. The functional used here is the one termed
Orsay-Paris~\cite{op}, which was shown to accurately reproduce the
static properties of pure and doped He clusters~\cite{fd}.  The need
to treat axially symmetric molecules implies moving from one to
two-dimensional equations.  The new routines have been extensively
tested against previously calculated spherically symmetric
systems. The minimization of energy is carried out by mapping the
density distribution on a grid of points and propagating it in
imaginary time, starting from a trial distribution.

The density functional also contains the interaction between the He and the
impurity molecule. The interaction, assumed pairwise, is treated as a
static external potential, since the molecules considered here are
expected to have negligible zero point motion. Existing potentials for
He-HCN~\cite{atki96}, He-HCCH~\cite{mosz95}, and He-OCS~\cite{higg99}
have been used without modifications.  The He-(HCN)$_2$ potential was
generated as the superposition of the potential due to two HCN
molecules whose centers of mass are separated by 4.44 ${\rm \AA}$ (the
equilibrium distance for the gas phase dimer \cite{juck88}). The
repulsive part of the He-HCCCN potential has been taken from
\cite{caba98}; the attractive part from the He-HCN and He-HCCH
potentials, using the concepts of distributed interaction and
transferability~\cite{bemi96}.  The He-HCCH \cite{mosz95} and
He-CH$_4$ \cite{buck85} interactions were used to generate the
potential between He and HCCCH$_3$, treating the latter molecule as
cylindrically symmetric. Full detail on all potentials used are
available from the authors, and will be published
separately~\cite{call00}.

Once the helium density profiles are calculated, the molecules are
assumed to rotate perpendicularly to their symmetry axis with angular
velocity $\omega$.  We assume that the He density adiabatically
follows this rotation, which allows us to calculate the
laboratory-frame time-dependent density at each point in the
liquid. This assumption is only valid if at each point the velocity of
the fluid, $v({\bf r})$, is less than a critical velocity, $v_c$. If
$v_c$ is taken to be the velocity of sound, this is true for all our
molecules, at the temperature of the droplet: 0.4 K.  A further
justification to our assumption is also the fact that no critical
value of angular momentum is experimentally observed for a wide class
of molecules (i.e. for a wide range of fluid velocities).

The second assumption that we make is that the He behaves entirely as
a superfluid undergoing irrotational flow.  The assumption that the
motion is irrotational implies that ${\bf v}({\bf r})$ can be written
as the gradient of a scalar potential: ${\bf v} =
-$\boldmath$\nabla$\unboldmath $\phi$ (the dependence of $\rho, {\bf
v},\phi$ on ${\bf r}$ will be implicit from now on), where $\phi$ is
known as the velocity potential.  These assumptions lead to the
following hydrodynamic equation for the velocity potential
\cite{miln96}:
\begin{equation}
\mbox{\boldmath$\nabla$} \cdot (\rho \mbox{\boldmath$\nabla$} \phi) =
\frac{\partial \rho}{\partial t}
= - (\mbox{\boldmath$\nabla$} \rho) \cdot ( \mbox{\boldmath$\omega$} \times
{\bf r} \,). \label{eq:hydro}
\end{equation}
The first equality is just the continuity equation, while the second
reflects the statement that the density is time-independent in the rotating
frame.  We select our axis system with $z$ along the symmetry axis of
the molecule, and assume that rotation takes place round the $x$ axis
with angular velocity $\mbox{\boldmath$\omega$} = \omega \, {\bf\hat{x}}$.
In order to better exploit the symmetry of the
problem, we have used elliptical coordinates $\xi,\theta,\varphi$,
where $x = f \, \sqrt{\xi^2-1} \sin (\theta) \cos(\varphi),\,\, y = f
\, \sqrt{\xi^2-1} \sin (\theta) \, \sin(\varphi)$, and $z = f \, \xi
\, \cos(\theta)$.  The surfaces of constant $\xi$ are ellipses of
rotation with foci at $z = \pm f$. Two such surfaces limit the region
where Eq.~(\ref{eq:hydro}) is solved. The inner boundary excludes the
volume occupied by the impurity, and is chosen as the largest ellipse
contained in the region where $\rho < 0.005 \rho_0$ ($\rho_0 = 0.0218
\AA^{-3}$ is the bulk liquid density).  Von Neumann boundary
conditions $\hat{n}\cdot\mbox{\boldmath$\nabla$}\phi =
-\hat{n}\cdot(\mbox{\boldmath$\omega$} \times {\bf r})$ insure that the
normal component of velocity matches the normal component of motion of
the boundary \cite{miln96}. For the outer boundary, any ellipse large
enough that the motion of the outside fluid is negligible can be
chosen (with Dirichlet boundary conditions $\phi = 0$).  These
boundary conditions result in a unique solution to the hydrodynamic
equations.  Other solutions exist if we do not require the fluid to be
irrotational, but it is known that these are higher in energy
\cite{miln96_k}, and will include any solutions that have some portion
(a ``normal component'' or a He ``snowball'') of the He density that
rigidly rotates with the molecule.

Given the solution, $\phi$, to the hydrodynamic equation, we can
calculate the kinetic energy, $K_{\rm h}$, in the motion of the fluid by
the following:
\begin{eqnarray}
K_{\rm h} &=& \frac{1}{2}\, I_{\rm h}\, \omega^2  \, =
\frac{1}{2} \, m_{\rm He} \, \int \, \rho \,
(\mbox{\boldmath$\nabla$} \phi) \cdot (\mbox{\boldmath$\nabla$} \phi) \, dV
\label{ekin0}
\\ K_{\rm h} &=& \frac{1}{2} \, m_{\rm He} \left[
- \int \, \phi \, \left( \frac{\partial \rho}{\partial t} \right) dV
+ \int \, \rho \, \phi \, (\mbox{\boldmath$\nabla$} \phi) \cdot
d{\bf S} \right].
\label{ekin}
\end{eqnarray}
Eq.~(\ref{ekin0}) follows directly from the definition of kinetic
energy; Eq.~(\ref{ekin}) is derived from Eq.~(\ref{ekin0}) using
standard vector identities and assuming that $\phi$ is a solution of
Eq.~(\ref{eq:hydro}).  $d{\bf S}$ is defined as positive when
pointing out of the region of the fluid.
$I_{\rm h}$ is the hydrodynamic contribution to
the moment of inertia for rotation about the $x$ axis, and $m_{\rm
He}$ is the atomic mass of helium.  Both $(\partial \rho / \partial
t)$ and $\phi$, are proportional to $\omega$, thus the above
definition of $I_{\rm h}$ is independent of $\omega$. The total
kinetic energy of rotation will include the contribution from the
molecule, $K_{\rm m} = \frac{1}{2}\, I_{\rm m}\, \omega^2$, where
$I_{\rm m}$ is the moment of inertia of the free molecule.  We can
also calculate the net angular momentum created by the motion of the
He fluid: ${\bf J_{\rm h}} = m_{\rm He} \, \int \rho \, {\bf r}
\times (- \mbox{\boldmath$\nabla$} \phi) \, dV$. By use of standard
vector identities and Eq.~(\ref{eq:hydro}), this definition can be
shown to lead to ${\bf J_{\rm h}} = I_{\rm h} \,
\mbox{\boldmath$\omega$}$. 
The total angular momentum is the sum of that of the rotating
molecule and the total moment of inetia the sum of the
moment of inertia of the molecule and that due to
hydrodynamic motion of the superfluid.  The local shape
of the velocity field ${\bf v} ({\bf r})$ can be rather complex due to
the presence of strong inhomogeneities in the density distribution.

We calculate $I_{\rm h}$ by solving the hydrodynamic equation,
Eq.~(\ref{eq:hydro}) for $\phi$, assuming unit angular velocity
rotation around the $x$ axis.  It is computationally convenient to
solve a slightly transformed version of Eq.~(\ref{eq:hydro}), where
the smoother function $\ln \rho ({\bf r})$ appears instead of
$\rho({\bf r})$:
\begin{equation}
\nabla^2 \phi + \left( {\bf \nabla} \ln \rho \right) \cdot \left(
{\bf \nabla} \phi  +  \hat{x} \times {\bf r} \right) = 0.
\label{eq:hydro2}
\end{equation}
Eq.~(\ref{eq:hydro2}) is solved, subject to the boundary conditions,
by converting it to a set of finite difference equations on a grid of
points in our elliptical coordinate system and using the {\it
Gauss-Seidel} relaxation method \cite{nr19}.  Both
Eq.~(\ref{ekin0}) and Eq.~(\ref{ekin}) are then evaluated by simple
numerical quadrature, and are found to give the same value of $I_{\rm
h}$ within a few percent. We also carefully tested the convergence of
$I_{\rm h}$ with grid size.

As an example of density distribution and velocity field,
Fig.~\ref{figure-ocs} shows our results for the OCS molecule in a
cluster of 300 He atoms. On the left we give the contour plot of the
He density near the molecule. One clearly sees the complex structure
which results from the tendency to have He atoms near local minima of
the impurity-He potential. The highest peak, at $(y,z)=(-3.6,-1.2)$,
corresponds to a ring of atoms perpendicular to the axis of the
molecule. The integral of the density within this structure gives 6.5
atoms, and indeed 7--8 is the number of He atoms one expects to fit
into such a ring by close-packing.  On the right side of the same
figure we plot the current density, $\rho \, {\bf v}$. We find that
most of the kinetic energy density, $\frac{1}{2}\rho \,{\bf v}^2$,
comes from the first solvation layer, the outer part of the cluster
giving a negligible effect.

In Table~\ref{tab1} our results are compared with existing
experimental values for several molecules in He nanodroplets.  There
is an overall good agreement between the predicted and observed
enhancements of the effective moment of inertia. From a quantitative
viewpoint, one notices that the predicted moments of inertia tend to
{\it overestimate} the experimental values. In the case of the
lightest rotors (HCN and HCCH) the large discrepancy suggests the
breakdown of the assumption of adiabatic following as recently
predicted~\cite{lee99}.  In that paper the importance of He exchange
is pointed out; it is also shown that the interplay of the rotational
constant with the potential anisotropy determines the extent to which
the anisotropic He solvation density can adiabatically follow the
rotation of the molecule. When the rotational constant of SF$_6$ is
arbitrarily increased in the calculation by a factor of 10, the He
density in the molecule-fixed frame becomes much more isotropic and
the solvation-induced $\Delta I$ decreases by a factor of
20~\cite{lee99}. We have recently obtained experimental evidence that
$\Delta I$ is larger for DCN than for HCN, which we believe to be
direct experimental evidence for this effect~\cite{conj99}. It is
interesting to note that for these light (i.e. fast spinning) rotors
the maximum of ${\bf v}({\bf r})$ approaches the bulk $^4$He sound
velocity.  

The overestimate of the moments of inertia for the other molecules
likely reflects the uncertainties in the calculated $\rho({\bf
r})$. We should remark here that while, by construction, the
Orsay-Paris functional prevents $\bar{\rho}$ (the density averaged
over an atomic volume) from becoming much larger than $\rho_0$, the
functional was not constructed to deal with density gradients as high
as those found in the first solvation layer.  We observed that small
changes in the form of the He density within the deep potential well
of those molecules produce significant variations of the predicted
moments of inertia, limiting the accuracy of the final results to 20\%
-- 30\%. This uncertainty does not affect the main result emerging
from Table~\ref{tab1} that the hydrodynamic contribution to the moment
of inertia of these systems, instead of being negligible, is rather
large and can explain the observed rotational constants.

One could object that the density values found at the minima of the
He-molecule interaction potential (e.g. $\approx~11~\rho_0$ for OCS)
are too high to be treated as those of a liquid, and should be
interpreted as localized He atoms rigidly rotating with the molecule;
it has been proposed that the He density distribution around the
OCS-He$_6$ supermolecule is only weakly anisotropic and thus can
rotate without generating a significant hydrodynamic
contribution~\cite{greb99}.  We have calculated the above density
distribution, and found that it is still strongly anisotropic, leading
to a hydrodynamic moment of inertia of over 400 ${\rm u \cdot
\AA}^2$.  When combined with the moment of inertia of the OCS-He$_6$
supermolecule, this gives a total effective moment of inertia of over
650 ${\rm u \cdot \AA}^2$, dramatically larger than the
experimental value (230 ${\rm u \cdot \AA}^2$).

In summary, the spatial dependence of the He density, which is caused
by the molecule-He interaction, results in a hydrodynamic contribution
to the moment of inertia more than an order of magnitude larger (in
the case of the heavier rotors) than that predicted for the rotation
of a reasonably sized ellipsoid in He of uniform bulk liquid
density. Furthermore, the present calculations suggest that the
effective moments of inertia of molecules in He nanodroplets (and
likely also bulk He) can be quantitatively predicted by assuming
irrotational flow of a spatially inhomogeneous superfluid.

We are pleased to acknowledge useful discussions and/or the sharing of
unpublished information with D.~Farrelly, Y.~Kwon, E.~Lee,
R.~E.~Miller, K.~Nauta, L. Pitaevskii, and K.~B.~Whaley.  The work was
supported by the National Science Foundation.

\bibliographystyle{prsty} 

\begin{thebibliography}{10}
\bibitem{miln96_k} Kelvin's minimum energy theorem. The proof in:
  \cite{miln96} is easily extended to the non-uniform density case.
\end{thebibliography}

\begin{table}[]
  \begin{center}
    \begin{tabular}{llllll}
& $I_{\rm m}$   \ \ \ \ \ \ \ \ & $I_{\rm eff}$ (exp.) \ \ \ &
$\Delta I$\ \ \ & $I_{\rm h}$ (calc.) \ & ref.
    \ \ \\  \hline
HCN & 11.39  & 14.04 & 2.65 & 5.47 & \cite[a]{nauta99a} \\ \hline
DCN & 14.0  & 16.9 & 2.9 & 5.6 & \cite{call00} \\ \hline
HCCH & 14.26 &  16.08 & 1.82 & 9.4 &\cite[c]{nauta99a} \\ \hline
HCCCH$_3$ &  59.14 & 224.0 & 164.8 & 190 & \cite{conj99}\\ \hline
OCS & 83.10 & 230.0 & 146.9 & 197 & \cite{greb98} \\ \hline
HCCCN & 111.1 & 330.7 & 219.6 & 226 & \cite{conj99} \\ \hline
(HCN)$_2$ & 289.5 & 872.5 & 583 & 619 &\cite[b]{nauta99a} \\ \hline
    \end{tabular}
    
\caption{Moments of inertia for the molecules studied in this
      work. Units are ${\rm u \cdot \AA}^2$. The quantities $I_{\rm m}$
      and $I_{\rm eff}$ are the observed moments of inertia when the
      molecule is free and dissolved in the cluster,
      respectively. Their difference, $\Delta I$, in the 4$^{th}$
      column is compared with the hydrodynamic moment of inertia,
      $I_{\rm h}$, of the present calculation.}

    \label{tab1}
  \end{center}
\end{table}

\begin{figure}

\caption{He density, $\rho$, (left) and He current density, $\rho \,
{\bf v}$, (right) distributions for a cluster of 300 He atoms, with
OCS in its center rotating counterclockwise. For the sake of showing
details, the highest He density peaks have been clipped (white areas),
and the dynamic range of the current has been compressed.}
\label{figure-ocs}
\end{figure}
\end{document}